\documentclass[%
 reprint,onecolumn,
superscriptaddress, 
12pt, 
nofootinbib,
 amsmath,amssymb,
floatfix,
]{revtex4-2}

\usepackage{graphicx}% Include figure files
\usepackage{dcolumn}% Align table columns on decimal point
\usepackage{bm}% bold math
\usepackage{cancel}% to cancel terms in equations
\usepackage{amsfonts}%bold math
\usepackage{xcolor}%to color some texts
\usepackage{tcolorbox}% for boxing texts

\newcommand{\beqa}{\begin{eqnarray}}
\newcommand{\eeqa}{\end{eqnarray}}
\newcommand{\nn}{\nonumber}
\DeclareMathOperator{\sech}{sech}

\begin{document}

\title[Article Title]{Co-rotating Vortices on Surfaces of Variable Negative Curvature: Hamiltonian Structure and Curvature-Induced Drift}
\author{
Gaurang Mangesh Joshi\,
\raisebox{-0.45ex}{\includegraphics[height=2.2ex]{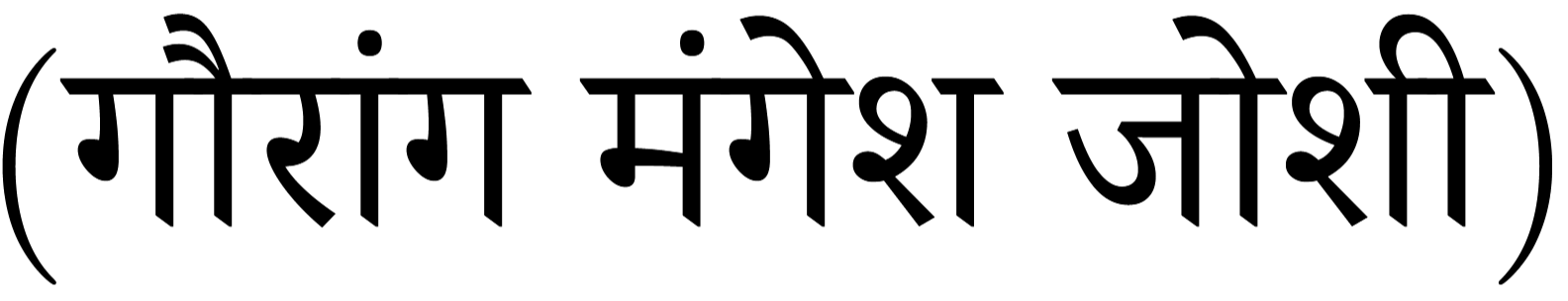}}
}
\affiliation{
Birla Institute of Technology and Science, Pilani, Hyderabad Campus,
Telangana 500078, India
}

\author{
Rickmoy Samanta\,
\raisebox{-0.45ex}{\includegraphics[height=2.2ex]{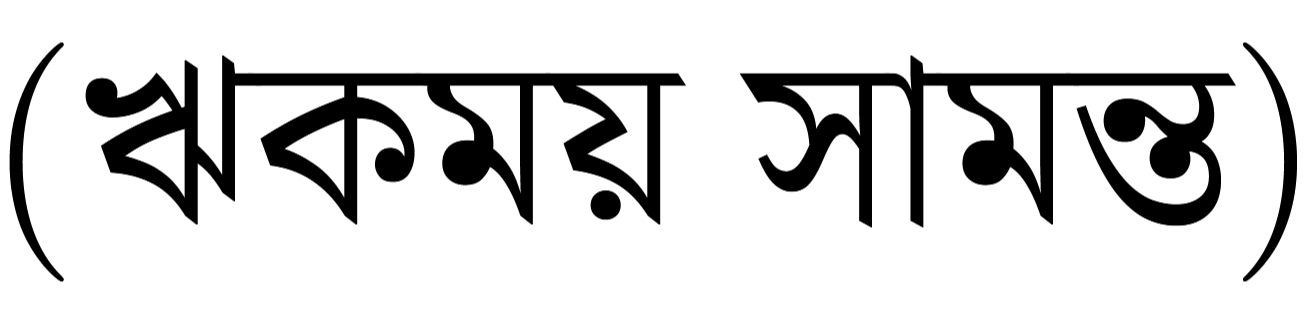}}
}
\affiliation{
Birla Institute of Technology and Science, Pilani, Hyderabad Campus,
Telangana 500078, India
}
\affiliation{
Indian Institute of Technology Kharagpur,
West Bengal 721302, India
}

\begin{abstract}
Vortices in fluids and superfluids are fundamental to phenomena ranging from Bose-Einstein condensates and superfluid films to neutron stars and
hydrodynamic micro-rotors, where background geometry often plays an important role. Curvature can induce vortex motion distinct from planar domains.
We study Hamiltonian vortex motion on a catenoid, a minimal surface of
variable negative curvature, and derive explicit equations of motion and 
conserved quantities for co-rotating vortex pairs. For
two identical vortices we find an exact analytic solution in which the
pair rotates rigidly at fixed latitude, with angular velocity
$\Omega=(\Gamma/16\pi)\,K'(V)/\sqrt{-K(V)}$, where $K(V)$ is the
Gaussian curvature. Thus the motion is governed by the curvature
gradient rather than the curvature itself. This  state is
linearly unstable, with growth rate $\lambda=\sqrt{3}|\Omega|$, in
agreement with numerical simulations. For generic co-rotating pairs,
conservation of the Hamiltonian and rotational momentum reduces the
nonlinear dynamics to a single quadrature, yielding bounded relative
oscillations together with a secular azimuthal drift. Simulations of the full equations confirm this and reveal the same
curvature-induced azimuthal drift in a localized many-vortex
cluster, motivating a broader theory of collective vortex drift on
curved surfaces.
\end{abstract}
\maketitle
%\tableofcontents

\section{Introduction}
Vortices are fundamental carriers of vorticity in a wide range of physical systems, from Bose--Einstein condensates and superfluid films to astrophysical fluids such as neutron star interiors, as well as engineered hydrodynamic rotors. A minimal theoretical framework for their motion is provided by point-vortex interactions, which offers one of the simplest Hamiltonian descriptions of two-dimensional rotating incompressible flow. In the plane, the motion is governed by the Kirchhoff--Routh function and inherits the Euclidean symmetries of the underlying domain, leading to conserved linear and angular momenta and logarithmic pair interactions \cite{lin1,lin2,aref,saffman}. Classical vortex pairs, dipoles, and clusters have therefore served as essential models for nonlinear vortex motion \cite{Tchieu2012,Lydon2022}.

On curved surfaces, geometry fundamentally alters vortex motion. Both the hydrodynamic Green function and the self-interaction, also known as the Robin function, acquire explicit geometric dependence. As a result, curvature acts as an effective field: it breaks translational invariance, modifies the phase-space structure, and generates secular drift  with no direct planar analogue.

Several complementary investigations have shaped the study of point vortices on curved manifolds. Vortex dynamics on spheres and other constant-curvature surfaces has shown how global topology and curvature modify relative equilibria, streamline topology, and integrability properties \cite{bg,kimok,kimura,newton1,newton2}. More general formulations on surfaces of revolution and closed curved manifolds have clarified the role of the Kirchhoff--Routh function, momentum maps, and curvature-dependent Robin terms \cite{hally,Koiller2009,boattok,boattod,Turner2010}. Related analytical developments include explicit constructions of path functions in multiply connected domains \cite{crowdymarshall}, studies of curvature-vortex coupling in surface flows \cite{voigt}, and recent work on vortex  dipoles on curved surfaces \cite{Gustafsson2022,khesin2024}. Recent studies of curved fluid membranes have further shown that geometry can organize hydrodynamic rotors, generate nontrivial streamline patterns, and collective cluster dynamics \cite{sam1,sam2,sam3}. These developments suggest that variable curvature is not merely a geometric background, but an active ingredient in the organization of vortex motion.

Curved-surface vortex dynamics is also relevant for quantum fluids. In thin superfluid films and quasi-two-dimensional Bose--Einstein condensates, point vortices provide an effective description of quantized vortex motion. Experiments and models have shown that trapping geometry, anisotropy, and curvature can strongly affect vortex trajectories, vortex-pair dynamics, vortex clusters, and turbulent states \cite{Neely2010,Freilich2010,Rooney2011,Goodman2015,white2012,White2014,Stagg2016,vsc,Caracanhas2022}. These systems motivate the search for analytically tractable geometries in which the consequences of curvature can be isolated and understood in closed form.

Despite these developments, the nonlinear dynamics of co-rotating vortex pairs on surfaces of variable negative curvature remains comparatively less explored. In particular, while vortex dipoles and collective motion on curved geometries have been investigated in recent work \cite{banthia2026,aswathy2026}, an explicit Hamiltonian reduction and analytic characterization of equal-strength co-rotating vortex pairs on the negatively curved surfaces has not been developed. The present work addresses this problem.

In this paper we focus on co-rotating point vortices on a catenoid. The catenoid is a minimal surface with tunable throat radius $a$ and nonuniform negative Gaussian curvature. Its axial symmetry yields a conserved momentum, while its conformal metric allows the Green function, Hamiltonian, symplectic form, and equations of motion to be written explicitly. This makes the catenoid a useful example for studying how curvature influences vortex motion beyond constant-curvature geometries.

We first examine the global Hamiltonian formulation for $N$ vortices on the catenoid, revisiting and extending the framework of Ref.~\cite{banthia2026}, where dipolar vortex motion was studied. In contrast, the present work focuses on co-rotating vortex pairs. The Hamiltonian consists of the pairwise Green-function interaction together with a curvature-induced self-interaction term. The corresponding symplectic form is weighted by the catenoid area element, and the rotational symmetry gives a momentum map. These two conserved quantities, the Hamiltonian $H$ and the momentum $J$, provide the main organizing structure for the two-vortex dynamics.

Our main analysis concerns two equal-strength co-rotating vortices. We show that the system is Liouville integrable: fixing $J$ eliminates the collective meridional coordinate, while fixing $H$ eliminates the relative azimuthal angle, leaving a single integral solution for the relative meridional separation. This reduction gives a complete analytic description of the nonlinear two-vortex motion and allows direct reconstruction of the mean azimuthal drift.

We then identify an exact symmetric solution in which two identical vortices remain antipodal at a fixed latitude and rotate rigidly around the catenoid. We show that the angular velocity of this configuration is controlled by the curvature gradient rather than by the curvature itself. In particular, if \(K(V)\) denotes the Gaussian curvature along the
meridional coordinate, the angular velocity can be expressed as a local
geometric combination proportional to
\(
K'(V)/\sqrt{-K(V)}
\).
Consequently, the rotation vanishes at the throat, where the curvature
is extremal and \(K'(V)=0\). The rotation rate instead reaches its
largest magnitude at finite distance from the throat and decays again
toward zero asymptotically. We also show that this symmetric co-rotating state is linearly unstable, with growth rate proportional to the angular velocity of rigid-rotation. Direct numerical simulations of the full two-vortex equations confirm both the instability rate and the reduced quadrature dynamics. Finally, we include a brief outlook on many-vortex dynamics. A representative simulation of a localized same-sign vortex cluster shows coherent azimuthal drift while the cluster remains compact. This indicates that the curvature-induced azimuthal drift found for vortex pairs persists in the many-vortex problem.\\ 
The paper is organized as follows. In Sec.~\ref{ham} we introduce the catenoid
geometry and derive the Hamiltonian formulation for $N$ point vortices,
including the Green function, symplectic form, equations of motion, and the
conserved angular momentum. In Sec.~\ref{2v} we specialize to two vortices and
rewrite the dynamics in collective and relative variables; the conserved
Hamiltonian and momentum are then used to reduce the two-vortex problem to a
single quadrature. In Sec.~\ref{symconfig} we identify an exact symmetric
co-rotating solution in which two identical vortices remain antipodal at fixed
latitude and rotate rigidly around the catenoid. We then analyze the linear
stability of this solution and verify the predicted instability rate
numerically in Sec.~\ref{numver1}. In Sec.~\ref{genrot} we study generic
equal-strength co-rotating pairs, showing that the reduced dynamics produces
bounded meridional oscillations together with a secular azimuthal drift. The
quadrature reduction and drift reconstruction are tested against full
numerical simulations in Sec.~\ref{numver2}. Finally, in
Sec.~\ref{clusteroutlook}, we give an illustration of many-vortex dynamics by
showing that a localized same-sign cluster remains compact while drifting
coherently around the catenoid. The Liouville integrability of the two-vortex system and
the associated Poisson structure are established explicitly in
Appendix~\ref{app1}.
\section{Model Setup}
\label{ham}
We begin by setting up the geometric and Hamiltonian ingredients needed for
the rest of the paper. The key point is that, on a curved surface, vortex
motion is affected not only by the interaction between different vortices but
also by the way the surface geometry changes from point to point. For the
catenoid these effects can be written explicitly, which allows us to derive a
closed dynamical system for vortices in the global coordinates $(v,u)$.  A detailed derivation of the Hamiltonian structure, including the conserved quantities on surfaces of variable negative curvature,
is given in Ref.~\cite{banthia2026}.

We consider a catenoid of throat radius $a>0$, parametrized by
\[
X(v,u) = \big( a\cosh(v/a)\cos u,\; a\cosh(v/a)\sin u,\; v\big),
\qquad u\in[0,2\pi),\; v\in\mathbb{R}.
\]
Here $u$ is the azimuthal angle around the symmetry axis, while $v$ measures
position along the meridional direction. The induced metric is
\[
g = \cosh^2\!\big(\tfrac{v}{a}\big)\,dv^2 
+ a^2\cosh^2\!\big(\tfrac{v}{a}\big)\,du^2
= \cosh^2\!\big(\tfrac{v}{a}\big)\,\big( dv^2 + a^2 du^2\big),
\]
with corresponding area element
\[
dA = a\cosh^2\!\big(\tfrac{v}{a}\big)\, dv\,du.
\]
The azimuthal periodicity in $u$ allows the hydrodynamic Green’s function to be written as
\[
G(v_i,u_i;v_j,u_j)
= \frac{1}{4\pi}\,
\log\!\Big(
\cosh\!\big(\tfrac{v_i-v_j}{a}\big) - \cos(u_i-u_j)
\Big).
\]
For notational convenience, we introduce
\[
F_{ij} := \cosh\!\big(\tfrac{v_i-v_j}{a}\big) - \cos(u_i-u_j),
\qquad
h(v) := \cosh\!\big(\tfrac{v}{a}\big).
\]
For a system of $N$ point vortices with circulations $\Gamma_i$ located at
$(v_i,u_i)$, the interaction Hamiltonian is
\begin{eqnarray}
H
&=&
\sum_{1\le i<j\le N}
\Gamma_i\Gamma_j\,G(v_i,u_i;v_j,u_j)
-\frac{1}{4\pi}\sum_{i=1}^N \Gamma_i^2\log h(v_i).
\label{hm}
\end{eqnarray}
The corresponding Hamilton equations are \cite{banthia2026}
\begin{eqnarray}
\Gamma_i a h^2(v_i)\,\dot v_i
&=&
\frac{\partial H}{\partial u_i},
\qquad
\Gamma_i a h^2(v_i)\,\dot u_i
=
-\frac{\partial H}{\partial v_i}.
\label{eq:hamilton_equations}
\end{eqnarray}
Differentiating the Green’s function gives
\[
\frac{\partial G}{\partial u_i}
=
\frac{1}{4\pi}\,
\frac{\sin(u_i-u_j)}{F_{ij}},
\qquad
\frac{\partial G}{\partial v_i}
=
\frac{1}{4\pi a}\,
\frac{\sinh\!\big(\tfrac{v_i-v_j}{a}\big)}{F_{ij}},
\]
while the curvature-induced self-interaction contributes
\[
\frac{\partial}{\partial v_i}
\Big(-\frac{1}{4\pi}\Gamma_i^2 \log h(v_i)\Big)
=
-\frac{1}{4\pi a}\Gamma_i^2\,\tanh\!\big(\tfrac{v_i}{a}\big).
\]
Substituting these expressions into Hamilton’s equations leads to the explicit dynamical system
\begin{equation}
\begin{aligned}
\dot v_i
&=
\frac{1}{4\pi a\,h^2(v_i)}
\sum_{j\neq i}
\Gamma_j\,
\frac{\sin(u_i-u_j)}{F_{ij}},
\\[1.2ex]
\dot u_i
&=
-\frac{1}{4\pi a^2\,h^2(v_i)}
\sum_{j\neq i}
\Gamma_j\,
\frac{\sinh\!\big(\tfrac{v_i-v_j}{a}\big)}{F_{ij}}
+
\frac{1}{4\pi a^2\,h^2(v_i)}
\Gamma_i\,\tanh\!\big(\tfrac{v_i}{a}\big).
\end{aligned}
\label{dyneq}
\end{equation}
These equations form the basis for all subsequent analysis. We now identify a second conserved quantity arising from symmetry; see Ref.~\cite{banthia2026}. The catenoid is invariant under rotations $u \mapsto u+\theta$, corresponding to a $U(1)$ action. The associated conserved quantity is obtained via the momentum map. Contracting the symplectic form with the generator $\partial/\partial u$ gives
\[
\iota_{\partial/\partial u}\omega
=
\sum_{i=1}^N \Gamma_i\, a\,h^2(v_i)\, dv_i.
\]
This must be an exact differential, implying
\[
\iota_{\partial/\partial u}\omega
=
d\!\left( \sum_{i=1}^N \Gamma_i S(v_i) \right),
\qquad
S'(v) = a\,h^2(v).
\]
Integrating,
\[
S(v)
=
\frac{a}{2}v
+
\frac{a^2}{4}\sinh\!\big(\tfrac{2v}{a}\big).
\]
Hence the conserved momentum is
\beqa
J
=
\sum_{i=1}^N \Gamma_i
\left(
\frac{a}{2}v_i
+
\frac{a^2}{4}\sinh\!\big(\tfrac{2v_i}{a}\big)
\right),
\qquad
\frac{dJ}{dt}=0.
\label{jdef}
\eeqa
Together with the conservation of the Hamiltonian $H$, these invariants provide important constraints on the dynamics and serve as useful diagnostics in the analysis that follows.

\section{Collective Variables and Quadrature Reduction for Two Vortices}
\label{2v}
To explore the geometric structure of the two--vortex motion on the catenoid, it is helpful to introduce the following collective and relative coordinates as follows
\begin{equation}
V = \frac{v_1 + v_2}{2}, 
\qquad 
\Delta v = v_1 - v_2,
\qquad
U = \frac{u_1 + u_2}{2}, 
\qquad 
\Delta u = u_1 - u_2,
\end{equation}
so that
\begin{equation}
v_1 = V + \frac{\Delta v}{2}, 
\qquad 
v_2 = V - \frac{\Delta v}{2}, 
\qquad
u_1 = U + \frac{\Delta u}{2}, 
\qquad 
u_2 = U - \frac{\Delta u}{2}.
\end{equation}
As before, we let $h(v)=\cosh(v/a)$ and define
\begin{equation}
h_1 = h(v_1), 
\qquad 
h_2 = h(v_2), 
\qquad 
F = \cosh\!\big(\tfrac{\Delta v}{a}\big) - \cos(\Delta u).\nn
\end{equation}
The equations of motion for two vortices of strengths $\Gamma_1$ and $\Gamma_2$ may be written as
\begin{align}
\dot v_1 &=
\frac{\Gamma_2}{4\pi a h_1^2}
\frac{\sin(\Delta u)}{F}, 
\qquad
\dot v_2 =
-\frac{\Gamma_1}{4\pi a h_2^2}
\frac{\sin(\Delta u)}{F}, \\
\dot u_1 &=
-\frac{\Gamma_2}{4\pi a^2 h_1^2}
\frac{\sinh(\Delta v/a)}{F}
+
\frac{\Gamma_1}{4\pi a^2 h_1^2}
\tanh\!\big(\tfrac{v_1}{a}\big), \\
\dot u_2 &=
+\frac{\Gamma_1}{4\pi a^2 h_2^2}
\frac{\sinh(\Delta v/a)}{F}
+
\frac{\Gamma_2}{4\pi a^2 h_2^2}
\tanh\!\big(\tfrac{v_2}{a}\big).
\end{align}
In terms of $(V,\Delta v,U,\Delta u)$  the above equations take the form
\begin{equation}
\dot V = \frac{\dot v_1 + \dot v_2}{2}, 
\qquad 
\dot{\Delta v} = \dot v_1 - \dot v_2,
\qquad
\dot U = \frac{\dot u_1 + \dot u_2}{2}, 
\qquad 
\dot{\Delta u} = \dot u_1 - \dot u_2,\nn
\end{equation}
and hence we obtain the closed system
\begin{align}
\dot V &=
\frac{\sin(\Delta u)}{8\pi a F}
\left(
\frac{\Gamma_2}{h_1^2}
-
\frac{\Gamma_1}{h_2^2}
\right), 
\label{eq:Vdot}\\[0.5ex]
\dot{\Delta v} &=
\frac{\sin(\Delta u)}{4\pi a F}
\left(
\frac{\Gamma_2}{h_1^2}
+
\frac{\Gamma_1}{h_2^2}
\right), 
\label{eq:dv}\\[0.5ex]
\dot U &=
\frac{1}{8\pi a^2}
\left[
-\frac{\sinh(\Delta v/a)}{F}
\left(
\frac{\Gamma_2}{h_1^2}
-
\frac{\Gamma_1}{h_2^2}
\right)
+
\frac{\Gamma_1\tanh(v_1/a)}{h_1^2}
+
\frac{\Gamma_2\tanh(v_2/a)}{h_2^2}
\right], 
\label{eq:Udot}\\[0.5ex]
\dot{\Delta u} &=
-\frac{\sinh(\Delta v/a)}{4\pi a^2 F}
\left(
\frac{\Gamma_2}{h_1^2}
+
\frac{\Gamma_1}{h_2^2}
\right)
+
\frac{1}{4\pi a^2}
\left[
\frac{\Gamma_1\tanh(v_1/a)}{h_1^2}
-
\frac{\Gamma_2\tanh(v_2/a)}{h_2^2}
\right].
\label{eq:du}
\end{align}
Equations \eqref{eq:Vdot}--\eqref{eq:du} provide an exact representation of the two--vortex dynamics in collective and relative variables. The terms proportional to $F^{-1}$ correspond to the geometric generalization of the planar Biot--Savart interaction, while the terms involving $\tanh(v/a)$ arise entirely from curvature-induced self-interaction.  Unlike the planar case, the conformal factor $h^2(v)$ prevents a complete decoupling between collective and relative degrees of freedom: the relative dynamics depends explicitly on $V$ through $h_1$ and $h_2$. This coupling is a direct manifestation of the nonuniform geometry of the catenoid. 
The system admits two conserved quantities. The two-vortex Hamiltonian may be written as
\begin{equation}
H
=
\frac{\Gamma_1\Gamma_2}{4\pi}
\log\!\Big(
\cosh\!\big(\tfrac{\Delta v}{a}\big) - \cos(\Delta u)
\Big)
-
\frac{1}{4\pi}
\left[
\Gamma_1^2 \log h(v_1)
+
\Gamma_2^2 \log h(v_2)
\right],
\end{equation}
with $v_{1,2}=V\pm \Delta v/2$. In addition, rotational invariance $u_i \mapsto u_i+\theta$ yields the conserved momentum
\begin{equation}
J
=
\Gamma_1 S(v_1)
+
\Gamma_2 S(v_2),
\qquad
S(v)
=
\frac{a}{2}v
+
\frac{a^2}{4}\sinh\!\big(\tfrac{2v}{a}\big),
\end{equation}
so that $dH/dt=0$ and $dJ/dt=0$. These invariants constrain the dynamics and, in principle, allow a reduction of the system to a single quadrature. The reduction proceeds as follows. Fixing the value of the momentum,
$J=J_0$, gives
\begin{equation}
J_0
=
\Gamma_1 S\!\left(V+\frac{\Delta v}{2}\right)
+
\Gamma_2 S\!\left(V-\frac{\Delta v}{2}\right).
\label{eq:Jconstraint}
\end{equation}
For fixed $J_0$ this equation determines the collective coordinate
$V$ implicitly as a function of the relative separation,
\begin{equation}
V=V(\Delta v;J_0),
\end{equation}
whenever the local inversion of \eqref{eq:Jconstraint} is valid. Thus
\[
v_1(\Delta v)=V(\Delta v;J_0)+\frac{\Delta v}{2},
\qquad
v_2(\Delta v)=V(\Delta v;J_0)-\frac{\Delta v}{2}.
\]
Next, fixing the energy $H=E$ gives
\begin{equation}
4\pi E
=
\Gamma_1\Gamma_2\log F
-
\Gamma_1^2\log h(v_1)
-
\Gamma_2^2\log h(v_2).
\end{equation}
For $\Gamma_1\Gamma_2\neq0$, this yields
\begin{equation}
F
=
\exp\!\left(\frac{4\pi E}{\Gamma_1\Gamma_2}\right)
h(v_1)^{\Gamma_1/\Gamma_2}
h(v_2)^{\Gamma_2/\Gamma_1}.
\label{eq:Fenergy}
\end{equation}
Since
\[
F=\cosh\!\left(\frac{\Delta v}{a}\right)-\cos(\Delta u),
\]
we obtain
\begin{equation}
\cos(\Delta u)
=
\cosh\!\left(\frac{\Delta v}{a}\right)
-
\exp\!\left(\frac{4\pi E}{\Gamma_1\Gamma_2}\right)
h(v_1)^{\Gamma_1/\Gamma_2}
h(v_2)^{\Gamma_2/\Gamma_1}.
\label{eq:cosdu}
\end{equation}
An important restriction follows from the energy relation. Since the right-hand side must lie in the interval $[-1,1]$, hence the allowed meridional separations are not arbitrary, but must satisfy
\begin{align}
-1
\leq
\cosh\!\left(\frac{\Delta v}{a}\right)
-
\exp\!\left(\frac{4\pi E}{\Gamma_1\Gamma_2}\right)
h(v_1)^{\Gamma_1/\Gamma_2}
h(v_2)^{\Gamma_2/\Gamma_1}
\leq
1.
\label{cond}
\end{align}
For convenience, let us also define
\begin{equation}
\mathcal C(\Delta v)
\equiv 
\cosh\!\left(\frac{\Delta v}{a}\right)
-
\exp\!\left(\frac{4\pi E}{\Gamma_1\Gamma_2}\right)
h(v_1)^{\Gamma_1/\Gamma_2}
h(v_2)^{\Gamma_2/\Gamma_1},
\end{equation}
where $v_1$ and $v_2$ are understood as functions of $\Delta v$ through
\eqref{eq:Jconstraint}. Then
\begin{equation}
\cos(\Delta u)=\mathcal C(\Delta v),
\qquad
\sin(\Delta u)
=
\epsilon\sqrt{1-\mathcal C^2(\Delta v)},
\label{eq:sindu}
\end{equation}
with $\epsilon=\pm1$ fixed by the initial condition and changing sign at turning points. Substituting \eqref{eq:sindu} into \eqref{eq:dv}, the relative meridional
motion obeys the single first-order equation
\begin{equation}
\dot{\Delta v}
=
\frac{\epsilon}{4\pi a\,F_E(\Delta v)}
\left[
\frac{\Gamma_2}{h^2(v_1)}
+
\frac{\Gamma_1}{h^2(v_2)}
\right]
\sqrt{1-\mathcal C^2(\Delta v)},
\label{eq:dvquadrature_ode}
\end{equation}
where
\begin{equation}
F_E(\Delta v)
=
\exp\!\left(\frac{4\pi E}{\Gamma_1\Gamma_2}\right)
h(v_1)^{\Gamma_1/\Gamma_2}
h(v_2)^{\Gamma_2/\Gamma_1}.
\end{equation}
Equivalently, the dynamics is reduced to the quadrature
\begin{equation}
t-t_0
=
\epsilon
\int_{\Delta v_0}^{\Delta v(t)}
\frac{
4\pi a\,F_E(\xi)\,d\xi
}{
\left[
\Gamma_2 h^{-2}(v_1(\xi))
+
\Gamma_1 h^{-2}(v_2(\xi))
\right]
\sqrt{1-\mathcal C^2(\xi)}
}.
\label{eq:quadrature}
\end{equation}
This is the desired one-dimensional reduction. The condition Eq.~(\ref{cond}) is precisely the requirement that the term $\sqrt{1-\mathcal C^2(\Delta v)}$ in the denominator of the reduced quadrature be real. The endpoints of the allowed interval occur when
$\cos(\Delta u)=\pm1$, corresponding respectively to
$\Delta u=0$ and $\Delta u=\pi$ modulo $2\pi$. These are the turning points
of the one-dimensional reduced motion in $\Delta v$. Once $\Delta v(t)$ is
obtained from \eqref{eq:quadrature}, the remaining variables follow by
reconstruction:
\begin{equation}
V(t)=V(\Delta v(t);J_0),
\qquad
\Delta u(t)=\arccos\mathcal C(\Delta v(t)),
\end{equation}
with the branch chosen consistently with \eqref{eq:sindu}. The mean
azimuthal coordinate is then obtained from
\begin{equation}
U(t)-U_0
=
\int_{t_0}^{t}
\dot U\big(V(s),\Delta v(s),\Delta u(s)\big)\,ds,
\end{equation}
where $\dot U$ is given in \eqref{eq:Udot}. Thus the two--vortex problem
on the catenoid is Liouville integrable: the conserved momentum removes
the collective meridional coordinate, the conserved Hamiltonian removes
the relative angle, and the remaining motion is governed by the single
quadrature \eqref{eq:quadrature}. A more explicit proof of Liouville integrability, including the symplectic structure and the involution of the conserved quantities, is presented in Appendix~\ref{app1}.
\subsection{Exact Rigidly Rotating Solution: Symmetric Two--Vortex Case}
\label{symconfig} 
\begin{figure}[t]
\centering
\includegraphics[width=\columnwidth]{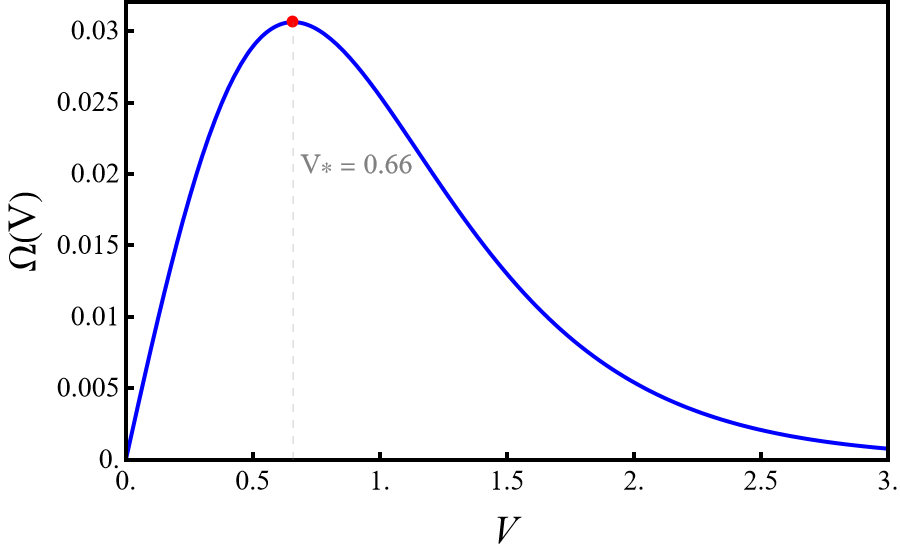}
\caption{
Angular velocity $\Omega(V)$ of the symmetric co--rotating vortex pair on a catenoid, given by 
$\Omega(V)=\frac{\Gamma}{4\pi a^2}\tanh(V/a)\sech^2(V/a)$, shown for $\Gamma=1$ and $a=1$. 
The curve exhibits a single global maximum at 
$V_*=\tfrac{a}{2}\ln(2+\sqrt{3})\approx 0.66\,a$, indicated by the red point. 
The angular velocity vanishes at the throat $V=0$, grows linearly for small $V$, and decays exponentially for $|V|\gg a$. 
}
\label{fig:omega_symmetric}
\end{figure}

We now identify an exact solution corresponding to a symmetric co--rotating configuration of two identical vortices, for which
\(
\Gamma_1=\Gamma_2=\Gamma>0.
\)
We consider the ansatz
\begin{equation}
v_1(t)=v_2(t)=V(t),
\qquad
\Delta u(t)=u_1(t)-u_2(t)=\pi.
\label{sym_ansatz}
\end{equation}
Under this ansatz, the relative meridional separation vanishes,
\[
\Delta v = 0,
\qquad
h_1=h_2=h(V),
\qquad
F=\cosh(0)-\cos\pi=2.
\]
Substituting into the meridional equations \eqref{eq:dv}, we obtain
\begin{equation}
\dot v_1=\dot v_2=0,
\end{equation}
so that
\begin{equation}
V(t)=V_0,
\end{equation}
and the vortices remain on a fixed latitude of the catenoid. The azimuthal equations \eqref{eq:du} reduce to
\begin{equation}
\dot u_1=\dot u_2
=
\frac{\Gamma}{4\pi a^2 h^2(V_0)}
\tanh\!\left(\frac{V_0}{a}\right),
\end{equation}
since the interaction term proportional to $\sinh(\Delta v/a)$ vanishes identically. Thus both vortices rotate with the same angular velocity
\begin{equation}
\Omega(V_0)
=
\frac{\Gamma}{4\pi a^2}
\tanh\!\left(\frac{V_0}{a}\right)
\sech^2\!\left(\frac{V_0}{a}\right).
\label{eq:omega_symmetric}
\end{equation}
It follows that
\begin{equation}
u_1(t)=u_1(0)+\Omega(V_0)t,
\qquad
u_2(t)=u_2(0)+\Omega(V_0)t,
\end{equation}
with the initial condition $u_1(0)-u_2(0)=\pi$. Consequently,
\begin{equation}
\Delta u(t)=\pi,
\end{equation}
and the antipodal separation is preserved exactly. The resulting motion is a rigid rotation:
\begin{equation}
v_1(t)=v_2(t)=V_0,
\qquad
u_1(t)=u_1(0)+\Omega(V_0)t,
\qquad
u_2(t)=u_1(t)-\pi.
\label{eq:symmetric_exact_solution}
\end{equation}
This describes a pair of vortices rotating uniformly along a circular latitude of the catenoid. The rotation is entirely curvature controlled: in the planar limit, or at the throat $V_0=0$, the angular velocity vanishes. For this solution, the conserved quantities simplify to
\begin{equation}
H
=
\frac{\Gamma^2}{4\pi}\log 2
-
\frac{\Gamma^2}{2\pi}\log h(V_0),
\end{equation}
and
\begin{equation}
J
=
2\Gamma S(V_0),
\qquad
S(V_0)
=
\frac{a}{2}V_0
+
\frac{a^2}{4}\sinh\!\left(\frac{2V_0}{a}\right).
\end{equation}
Thus the conserved momentum fixes the latitude $V_0$, while the Hamiltonian determines the corresponding energy of the rigidly rotating pair. The dependence of the rotation rate on latitude admits a direct geometric interpretation. The Gaussian curvature of the catenoid is
\begin{equation}
K(V)=-\frac{1}{a^2\cosh^4(V/a)},
\end{equation}
with derivative
\begin{equation}
K'(V)=\frac{4}{a^3}\frac{\sinh(V/a)}{\cosh^5(V/a)}.
\end{equation}
Using these expressions, the angular velocity \eqref{eq:omega_symmetric} can be rewritten as
\begin{equation}
\Omega(V)=\frac{\Gamma}{16\pi}\,\frac{K'(V)}{\sqrt{-K(V)}},
\end{equation}
showing that the rigid rotation is controlled by the curvature gradient rather than by the curvature itself. The maximum rotation follows from
\begin{equation}
\frac{d\Omega}{dV}
=
\frac{\Gamma}{4\pi a^3}
\sech^2\!\left(\frac{V}{a}\right)
\left[1-3\tanh^2\!\left(\frac{V}{a}\right)\right],
\end{equation}
so that the nontrivial extrema occur at
\begin{equation}
\tanh\!\left(\frac{V_*}{a}\right)=\pm\frac{1}{\sqrt{3}}.
\end{equation}
In particular, $\Omega(V)$ vanishes at the throat $V=0$, where $K'(0)=0$, and changes sign across it. The location of the extrema can be obtained explicitly. From the condition
\(
\tanh(V_*/a)=\pm 1/\sqrt{3},
\)
we find
\begin{equation}
V_*=\pm a\,\operatorname{arctanh}\!\left(\frac{1}{\sqrt{3}}\right)
=\pm \frac{a}{2}\ln\!\left(\frac{\sqrt{3}+1}{\sqrt{3}-1}\right)
=\pm \frac{a}{2}\ln(2+\sqrt{3}).
\end{equation}
Thus the maximal magnitude of the angular velocity occurs at a finite distance from the throat, symmetrically located at $V=\pm V_*$. In particular, the rotation is not strongest at the point of maximal curvature ($V=0$), but rather where the curvature varies most rapidly. This provides a direct geometric characterization of the rigidly rotating state: the dynamics is governed by the spatial gradient of curvature, with the extrema marking the maximal curvature-induced rotation. In the asymptotic region $|V|\gg a$, we find
\begin{equation}
\Omega(V)
\sim
\frac{\Gamma}{\pi a^2}\,
\mathrm{sgn}(V)\,e^{-2|V|/a},
\end{equation}
reflecting the exponential decay of both the curvature and its gradient. The corresponding angular velocity profile $\Omega(V)$ is shown in Fig.~\ref{fig:omega_symmetric}, which illustrates the vanishing of the rotation rate at the throat, its growth to a finite maximum at intermediate latitude, and its exponential decay at large $|V|$.
\subsection{Linear Stability of the Symmetric Rigidly Rotating State}
\begin{figure*}[t]
\centering
\includegraphics[width=0.95\textwidth]{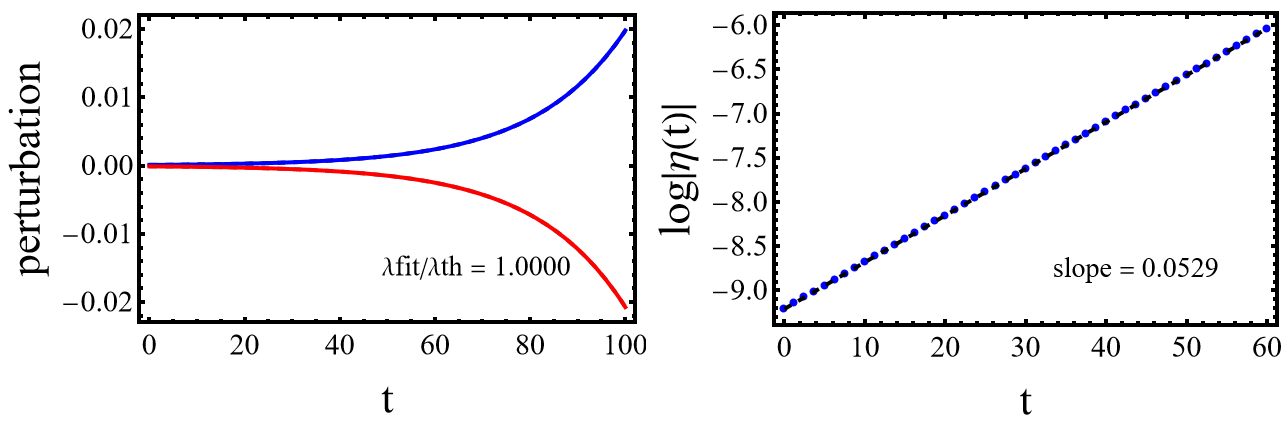}
\caption{
Linear stability tests of the symmetric co--rotating two--vortex configuration on the catenoid. 
(a) Time evolution of small perturbations $\eta(t)=\Delta v(t)$ and $\phi(t)=\Delta u(t)-\pi$, showing the growth predicted by the linearized equations. 
(b) Logarithmic plot of $|\eta(t)|$, whose fitted slope agrees with the theoretical instability rate $\lambda=\sqrt{3}\,|\Omega(V_0)|$ with $\delta J\lesssim 10^{-14}, \delta H \lesssim 10^{-14}$ through the time interval.}
\label{fig:linstab}
\end{figure*}
We now examine the linear stability of the symmetric antipodal solution. For identical co--rotating vortices, $\Gamma_1=\Gamma_2=\Gamma>0$, we perturb the rigidly rotating state according to
\begin{equation}
V(t)=V_0+\delta V(t),
\qquad
\Delta v(t)=\eta(t),
\qquad
\Delta u(t)=\pi+\phi(t),\nn
\end{equation}
where $|\delta V|$, $|\eta|$, and $|\phi|$ are small. To linear order,
\begin{equation}
\sin(\Delta u)=\sin(\pi+\phi)\simeq -\phi,
\qquad
F=\cosh(\eta/a)-\cos(\pi+\phi)\simeq 2.\nn
\end{equation}
Expanding Eqs.~\eqref{eq:Vdot}, \eqref{eq:dv}, and \eqref{eq:du} about the antipodal solution gives
\begin{equation}
\dot{\delta V}=0,
\qquad
\dot\eta
=
-\frac{\Gamma}{4\pi a}
\sech^2\!\left(\frac{V_0}{a}\right)\phi,
\qquad
\dot\phi
=
-\frac{3\Gamma}{4\pi a^3}
\sech^2\!\left(\frac{V_0}{a}\right)
\tanh^2\!\left(\frac{V_0}{a}\right)\eta .
\end{equation}
Thus the collective meridional perturbation is neutral at linear order, while the relative perturbations are coupled. Eliminating $\phi$ gives
\begin{equation}
\ddot\eta=\lambda^2\eta,
\end{equation}
with
\begin{equation}
\lambda^2
=
\frac{3\Gamma^2}{16\pi^2a^4}
\sech^4\!\left(\frac{V_0}{a}\right)
\tanh^2\!\left(\frac{V_0}{a}\right),
\end{equation}
so that
\begin{equation}
\lambda
=
\frac{\sqrt{3}\Gamma}{4\pi a^2}
\sech^2\!\left(\frac{V_0}{a}\right)
\left|
\tanh\!\left(\frac{V_0}{a}\right)
\right|.
\end{equation}
In terms of the rigid-rotation frequency \eqref{eq:omega_symmetric}, this takes the compact form
\begin{equation}
\lambda=\sqrt{3}\,|\Omega(V_0)|.
\end{equation}
Since $\lambda^2>0$ for $V_0\neq0$, the linearized dynamics admits an exponentially growing mode, implying that the symmetric co--rotating state is linearly unstable away from the throat. At $V_0=0$, one has $\Omega=\lambda=0$, and the instability becomes marginal at linear order.

\subsection{Numerical verification of the linear instability}
\label{numver1}
We now verify the above linear stability predictions by direct numerical integration of the full two--vortex equations of motion on the catenoid. Throughout this section we consider two identical vortices with $\Gamma_1=\Gamma_2=\Gamma>0$ and fix the geometric parameter $a=1$ without loss of generality. The symmetric rigidly rotating state is initialized at a prescribed latitude $V_0>0$, with small perturbations chosen to excite the unstable eigenmode of the linearized system. Specifically, we take
\begin{equation}
V(0)=V_0,\qquad
\Delta v(0)=\eta_0,\qquad
\Delta u(0)=\pi+\phi_0,
\end{equation}
where $|\eta_0|\ll1$. To isolate the growing mode, the initial angular perturbation is chosen according to the linear eigenvector relation
\begin{equation}
\phi_0
=
-\frac{\lambda}{A}\,\eta_0
=
-\frac{\sqrt{3}}{a}\tanh\!\left(\frac{V_0}{a}\right)\eta_0,
\label{choice}
\end{equation}
with
\(
A=\frac{\Gamma}{4\pi a}\sech^2(V_0/a).
\)
This choice follows directly from the structure of the linearized system. Writing the perturbation equations in the form
\begin{equation}
\dot\eta=-A\phi,
\qquad
\dot\phi=-B\eta,\nn
\end{equation}
where
\begin{equation}
B=
\frac{3\Gamma}{4\pi a^3}
\sech^2\!\left(\frac{V_0}{a}\right)
\tanh^2\!\left(\frac{V_0}{a}\right),\nn
\end{equation}
the perturbations evolve according to
\begin{equation}
\frac{d}{dt}
\begin{pmatrix}
\eta\\
\phi
\end{pmatrix}
=
\begin{pmatrix}
0 & -A\\
-B & 0
\end{pmatrix}
\begin{pmatrix}
\eta\\
\phi
\end{pmatrix}.\nn
\label{eq:matrixlin}
\end{equation}
The eigenvalues of the linearized matrix satisfy
\begin{equation}
\det
\begin{pmatrix}
-\mu & -A\\
-B & -\mu
\end{pmatrix}
=0,\nn
\end{equation}
which gives
\begin{equation}
\mu_\pm=\pm\lambda,
\qquad
\lambda=\sqrt{AB}.\nn
\end{equation}
For the unstable mode $\mu=+\lambda$, the associated eigenvector satisfies
\begin{equation}
\lambda\eta=-A\phi,
\end{equation}
or equivalently
\begin{equation}
\phi=-\frac{\lambda}{A}\eta
\end{equation}
motivating the choice of the initial perturbation dictated by Eq.~\ref{choice} and the numerical evolution thus isolates the exponentially growing instability,
\begin{equation}
\eta(t)\propto e^{\lambda t},
\qquad
\phi(t)\propto e^{\lambda t},
\end{equation}
allowing a direct comparison between the nonlinear numerical solution and the analytic growth rate predicted by the stability analysis.
In terms of the original vortex coordinates, this corresponds to
\begin{equation}
v_1(0)=V_0+\frac{\eta_0}{2},\qquad
v_2(0)=V_0-\frac{\eta_0}{2},\nn
\end{equation}
\begin{equation}
u_1(0)=\frac{\pi+\phi_0}{2},\qquad
u_2(0)=-\frac{\pi+\phi_0}{2}.\nn
\end{equation}
The equations of motion are then integrated numerically over a time interval $t\in[0,t_f]$, with $t_f$ chosen sufficiently large to resolve the exponential growth regime while remaining within the validity of the linear approximation. The full nonlinear equations were integrated numerically using Mathematica with adaptive-step explicit Runge--Kutta methods and default adaptive tolerance controls. Numerical stability and accuracy were monitored by tracking the conserved Hamiltonian $H$ and rotational momentum $J$ throughout the evolution. In the representative simulations considered here, the deviations
\[
\delta H(t)=H(t)-H(0),
\qquad
\delta J(t)=J(t)-J(0),
\]
remain small over the integration interval, with typical magnitudes
\[
|\delta J|\lesssim 10^{-14},
\qquad
|\delta H|\lesssim 10^{-14}.
\]
This provides an internal consistency check on the numerical integration. At each time step we extract the relative variables
\begin{equation}
\eta(t)=v_1(t)-v_2(t),
\qquad
\phi(t)=\Delta u(t)-\pi,
\end{equation}
and compare them with the linear predictions
\begin{equation}
\eta_{\rm lin}(t)=\eta_0 e^{\lambda t},
\qquad
\phi_{\rm lin}(t)=\phi_0 e^{\lambda t}.
\end{equation}

Figure~\ref{fig:linstab}(a) shows the time evolution of $\eta(t)$ and $\phi(t)$ obtained from the full nonlinear equations together with the predictions of the linearized theory, showing good agreement. In Fig.~\ref{fig:linstab}(b) we plot $\log|\eta(t)|$, which displays an approximately linear dependence on time over the early-time interval where the perturbation remains within the linear regime, confirming exponential growth. The numerical growth rate $\lambda_{\rm fit}$ is extracted from a linear regression of $\log|\eta(t)|$ over this interval and compared directly with the analytic prediction $\lambda$. For the parameters used in Fig.~\ref{fig:linstab}, we obtain
\[
\frac{|\lambda_{\rm fit}-\lambda|}{|\lambda|}
<10^{-3},
\]
demonstrating good agreement between the nonlinear evolution and the linear stability analysis. However, because the reduced system conserves both energy and angular momentum, the nonlinear dynamics remains globally bounded: perturbations grow exponentially while remaining confined to invariant energy and rotational-momentum contours.
\section{Generic Co--Rotating Vortex Pairs}
\label{genrot}
\begin{figure*}[t]
\centering

%================ TOP ROW =================%
\begin{minipage}{0.48\textwidth}
\centering
\includegraphics[width=\linewidth]{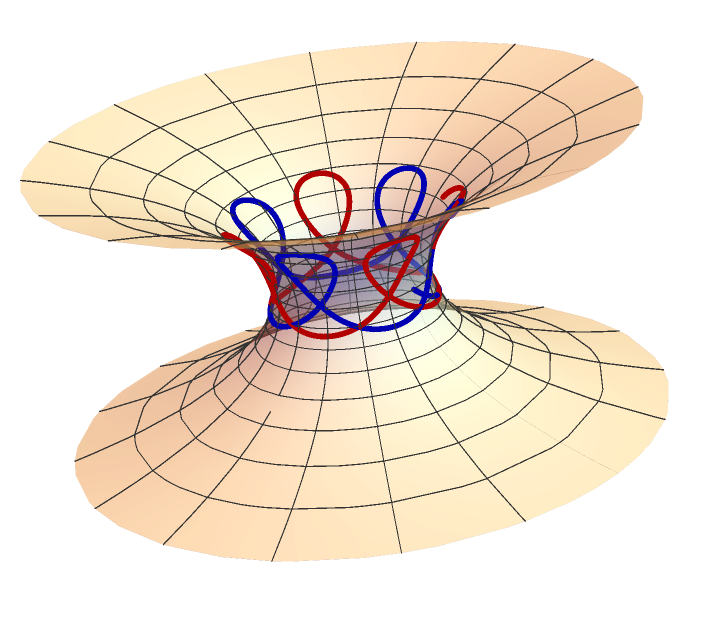}
\end{minipage}
\hfill
\begin{minipage}{0.48\textwidth}
\centering
\includegraphics[width=\linewidth]{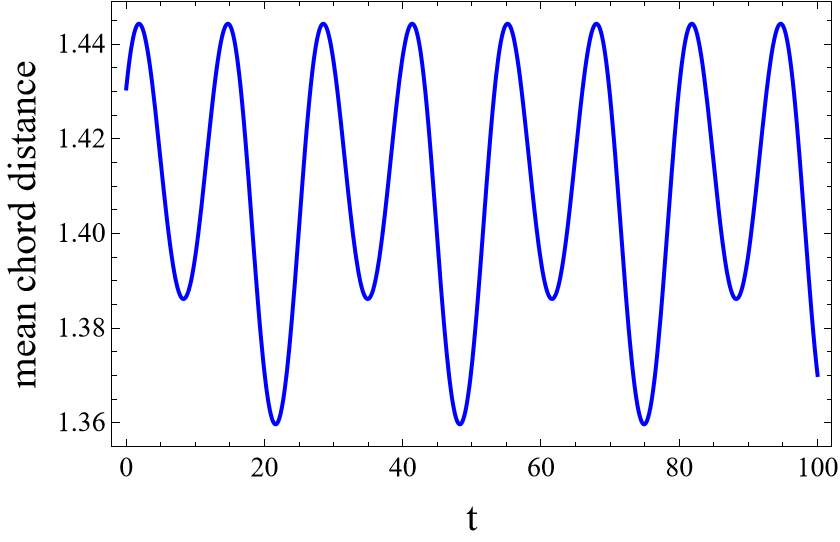}
\end{minipage}

\vspace{0.4cm}

%================ BOTTOM ROW =================%
\begin{minipage}{0.8\textwidth}
\centering
\includegraphics[width=\linewidth]{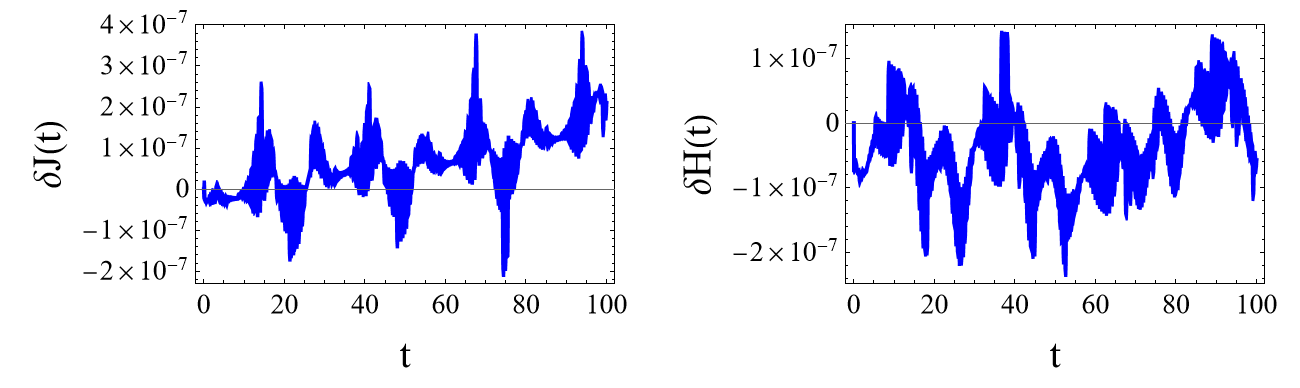}
\end{minipage}

\caption{
Dynamics of a generic equal--strength co--rotating vortex pair on the catenoid. 
Top left: three--dimensional trajectories on the embedded surface, showing quasi-periodic motion with azimuthal drift and bounded meridional oscillations. 
Top right: mean Euclidean chord distance between the vortices, exhibiting bounded oscillations consistent with integrable dynamics. 
Bottom: conservation of the invariants, showing the deviations $\delta J(t)$ and $\delta H(t)$, which remain at the level of numerical precision throughout the evolution. 
Together, these diagnostics confirm both the accuracy of the numerical integration and the constrained, quasi-periodic nature of the co--rotating motion.
}
\label{fig:generic_corotating_dynamics}
\end{figure*}
\begin{figure}[t]
\centering
\includegraphics[width=\columnwidth]{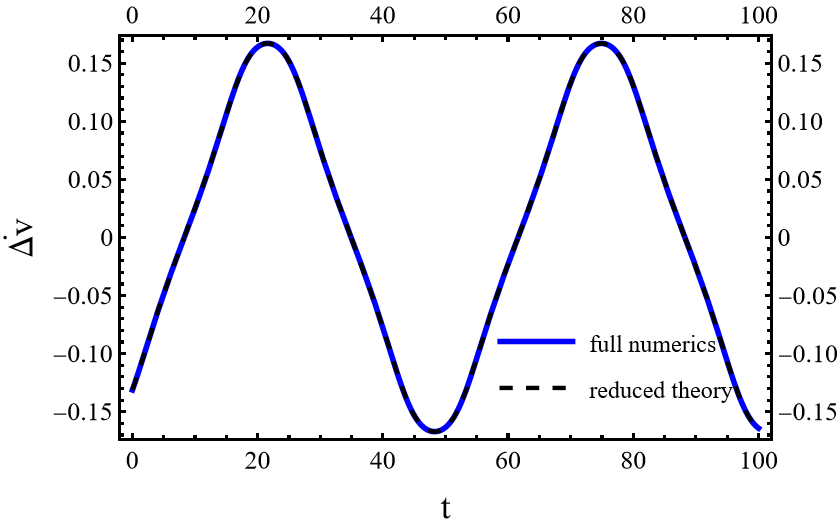}
\caption{
Comparison of the full numerical dynamics and the reduced analytic theory for a generic equal--strength co--rotating vortex pair on the catenoid. The solid curve shows the numerically computed $\dot{\Delta v}(t)$, while the dashed curve corresponds to the analytic prediction obtained from the reduced quadrature using the conserved quantities $H$ and $J$. 
}
\label{fig:thvsnum}
\end{figure}
\begin{figure*}[t]
\centering
\includegraphics[width=0.95\textwidth]{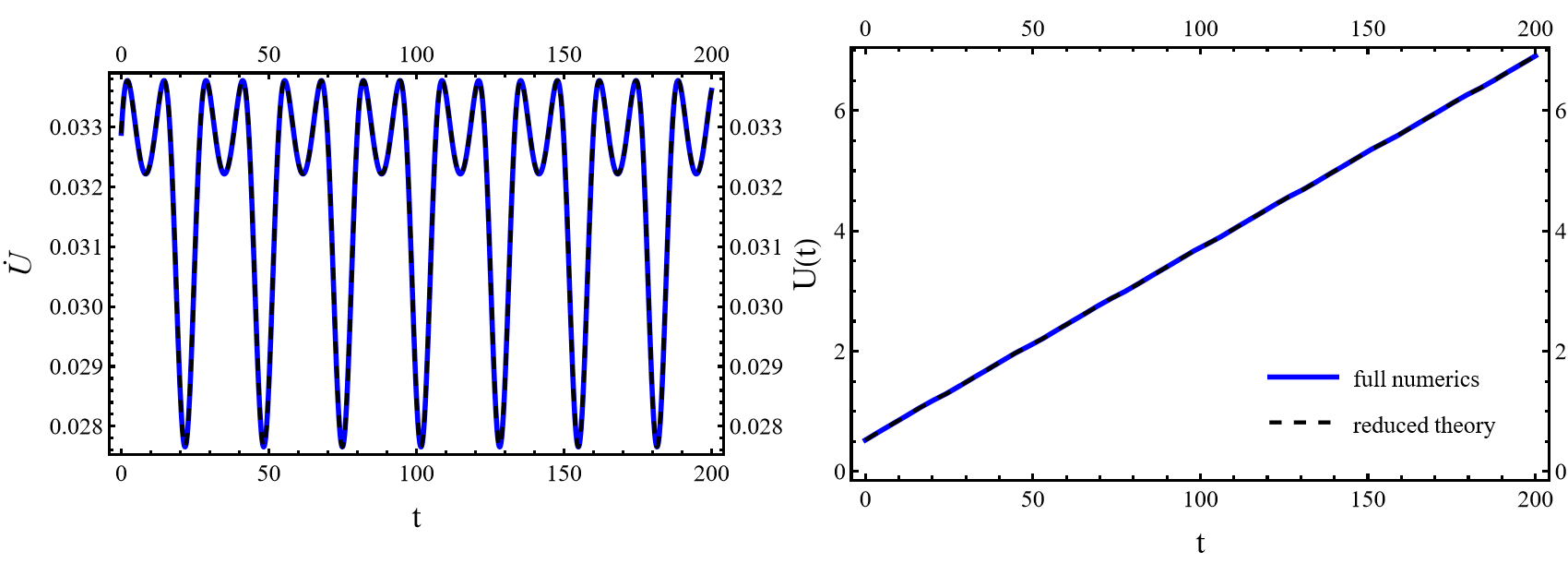}
\caption{
Verification of the reduced theory for the mean azimuthal drift of a generic equal--strength co--rotating vortex pair. 
Left: instantaneous drift rate $\dot U(t)$ obtained from the full numerical solution (solid blue) compared with the reduced prediction $\mathcal{U}(\Delta v(t);H,J)$ (black dashed). 
Right: reconstruction of the mean azimuthal coordinate $U(t)$ by integrating the reduced drift equation, showing good agreement with the full numerical evolution. 
}
\label{fig:uver}
\end{figure*}
We now develop an analytic description of a generic co--rotating pair of identical vortices on the catenoid, with
\(
\Gamma_1=\Gamma_2=\Gamma>0.
\)
In contrast to the symmetric antipodal configuration, the dynamics is fully two--dimensional in the relative variables and exhibits a combination of meridional oscillation and azimuthal drift. It is convenient to work in the collective and relative variables introduced in previous section,
\[
V=\frac{v_1+v_2}{2},\qquad
\Delta v=v_1-v_2,\qquad
U=\frac{u_1+u_2}{2},\qquad
\Delta u=u_1-u_2,
\]
so that
\(
v_{1,2}=V\pm \tfrac{\Delta v}{2}.
\)
For equal strengths, the Hamiltonian reduces to
\begin{equation}
H
=
\frac{\Gamma^2}{4\pi}\log\!\Big(
\cosh\!\big(\tfrac{\Delta v}{a}\big)-\cos(\Delta u)
\Big)
-
\frac{\Gamma^2}{4\pi}
\Big[
\log h(v_1)+\log h(v_2)
\Big],
\end{equation}
while the conserved momentum simplifies to the explicit form
\begin{equation}
J
=
\Gamma\Big(S(v_1)+S(v_2)\Big)
=
\Gamma\left[
a V
+
\frac{a^2}{2}\sinh\!\left(\frac{2V}{a}\right)
\cosh\!\left(\frac{\Delta v}{a}\right)
\right].
\label{eq:J_equal}
\end{equation}
Equation \eqref{eq:J_equal} provides an explicit algebraic relation between the collective coordinate $V$ and the relative separation $\Delta v$ at fixed $J=J_0$. Thus, unlike the general unequal--strength case, the collective meridional coordinate can be eliminated exactly without inversion ambiguities, yielding
\[
V=V(\Delta v;J_0).
\]
Substituting this relation into the Hamiltonian, one obtains an effective one--degree--of--freedom description. The energy constraint $H=E$ gives
\begin{equation}
\cos(\Delta u)
=
\cosh\!\left(\frac{\Delta v}{a}\right)
-
C_E\,h(v_1)h(v_2),
\qquad
C_E=\exp\!\left(\frac{4\pi E}{\Gamma^2}\right),
\label{eq:cosdu_equal}
\end{equation}
so that the relative angle is determined algebraically by $\Delta v$. Writing
\[
\sin(\Delta u)
=
\epsilon\sqrt{1-\mathcal C^2(\Delta v)},
\qquad
\mathcal C(\Delta v)=\cos(\Delta u),
\]
the meridional dynamics reduces to a single first--order equation
\begin{equation}
\dot{\Delta v}
=
\frac{\epsilon\Gamma}{4\pi a}
\frac{
h^{-2}(v_1)+h^{-2}(v_2)
}{
F_E(\Delta v)
}
\sqrt{1-\mathcal C^2(\Delta v)},
\label{eq:dv_equal}
\end{equation}
where
\(
F_E(\Delta v)=C_E\,h(v_1)h(v_2).
\)
This defines a one--dimensional integrable flow with turning points determined by $|\mathcal C(\Delta v)|\le1$. A key feature of the equal--strength case is the emergence of a nontrivial drift of the collective azimuthal coordinate. Averaging the individual azimuthal velocities yields
\begin{equation}
\dot U
=
\frac{\Gamma}{8\pi a^2}
\left[
-\frac{\sinh(\Delta v/a)}{F}
\Big(
\frac{1}{h^2(v_1)}-\frac{1}{h^2(v_2)}
\Big)
+
\frac{\tanh(v_1/a)}{h^2(v_1)}
+
\frac{\tanh(v_2/a)}{h^2(v_2)}
\right],
\label{eq:Udrift_equal}
\end{equation}
which is generically nonzero. This term represents a curvature--induced drift that has no analogue in the planar problem. Using \eqref{eq:cosdu_equal}, the chord distance \footnote{The geometric content of the motion is most transparent in the standard $\mathbb{R}^3$ embedding of the catenoid,
\[
X(u,v)=\big(a\cosh(v/a)\cos u,\;a\cosh(v/a)\sin u,\;v\big).
\]
The instantaneous chord distance between the vortices is then
\begin{equation}
d_{\mathbb R^3}^2
=
a^2\Big[
h^2(v_1)+h^2(v_2)
-2h(v_1)h(v_2)\cos(\Delta u)
\Big]
+
(\Delta v)^2,
\label{eq:chord_equal}
\end{equation}
which provides a natural geometric measure of the pair separation.}  can be expressed entirely in terms of $\Delta v$ and the conserved quantities.

Equations \eqref{eq:dv_equal} and \eqref{eq:Udrift_equal} show that the generic co--rotating motion consists of an oscillatory evolution in the relative coordinate $\Delta v(t)$, coupled to a secular azimuthal drift $U(t)$.
Thus, in contrast to the rigidly rotating symmetric state, the generic co--rotating pair exhibits a mixed dynamics combining oscillations in separation with curvature--induced motion along the azimuthal direction. The resulting dynamics is illustrated in Fig.~\ref{fig:generic_corotating_dynamics}. 
The vortex trajectories exhibit bounded meridional oscillations combined with a persistent azimuthal drift, while the inter-vortex separation remains oscillatory and finite.  The conservation of $H$ and $J$ is maintained to numerical precision.
\subsection{Verification of the reduced quadrature dynamics and azimuthal drift}
\label{numver2}
To validate the analytic reduction developed above, we compare the full numerical evolution of a generic equal--strength co--rotating pair with the reduced one--dimensional theory implied by the conserved quantities. Throughout this test we consider $\Gamma_1=\Gamma_2=\Gamma$ and fix parameters $a=1$ and $\Gamma=1$.
The initial conditions are taken to be
\begin{equation}
(u_1(0),v_1(0))=(0,0),
\qquad
(u_2(0),v_2(0))=\left(\frac{\pi}{3},\frac{\pi}{4}\right),\nn
\end{equation}
corresponding to
\begin{equation}
V_0=\frac{v_1(0)+v_2(0)}{2},
\quad
\Delta v_0=v_1(0)-v_2(0),
\quad
\Delta u_0=u_1(0)-u_2(0).\nn
\end{equation}
The full dynamics is obtained by direct numerical integration of the equations of motion, yielding $u_i(t)$ and $v_i(t)$, from which we extract the relative variables
\begin{equation}
\Delta v(t)=v_1(t)-v_2(t),
\qquad
\Delta u(t)=u_1(t)-u_2(t).\nn
\end{equation}
The reduced theory follows from the conservation of the Hamiltonian and momentum,
\begin{equation}
H
=
\frac{\Gamma^2}{4\pi}\log\!\big(\cosh(\Delta v/a)-\cos(\Delta u)\big)
-
\frac{\Gamma^2}{4\pi}\big[\log h(v_1)+\log h(v_2)\big],\nn
\end{equation}
\begin{equation}
J
=
\Gamma\left(aV+\frac{a^2}{2}\sinh\!\left(\frac{2V}{a}\right)\cosh\!\left(\frac{\Delta v}{a}\right)\right),\nn
\end{equation}
which determine $V=V(\Delta v)$ and the effective interaction factor
\begin{equation}
F_E(\Delta v)
=
\exp\!\left(\frac{4\pi H_0}{\Gamma^2}\right)
h(v_1)h(v_2).\nn
\end{equation}
Eliminating $\Delta u$ using
\begin{equation}
\cos(\Delta u)
=
\cosh\!\left(\frac{\Delta v}{a}\right)
-
F_E(\Delta v),\nn
\end{equation}
the relative dynamics reduces to the first--order equation
\begin{equation}
\dot{\Delta v}
=
\frac{\Gamma}{4\pi a\,F_E(\Delta v)}
\left[
\frac{1}{h^2(v_1)}+\frac{1}{h^2(v_2)}
\right]
\sin(\Delta u),
\end{equation}
with the branch of $\sin(\Delta u)$ fixed dynamically. In practice, $V(\Delta v)$ is obtained by solving $J(V,\Delta v)=J_0$, and the reduced right--hand side is evaluated along the trajectory using the instantaneous sign of $\sin(\Delta u(t))$. We first compare $\dot{\Delta v}(t)$ computed from the full numerical solution with the analytic prediction. As shown in Fig.~\ref{fig:thvsnum}, the two are indistinguishable over the entire time interval, demonstrating that the quadrature reduction exactly captures the dynamics, including the correct branch structure across turning points. This provides a verification of the Liouville integrability of the system, see also Appendix \ref{app1} for a mathematical derivation. A key consequence of the reduction is that the mean azimuthal coordinate
\begin{equation}
U(t)=\frac{u_1(t)+u_2(t)}{2}
\end{equation}
is not an independent dynamical variable, but is completely determined by $\Delta v(t)$. Its evolution is governed by
\begin{equation}
\dot U
=
\frac{\Gamma}{8\pi a^2}
\left[
-\frac{\sinh(\Delta v/a)}{F}\!\left(\frac{1}{\cosh^2(v_1/a)}-\frac{1}{\cosh^2(v_2/a)}\right)
+\frac{\tanh(v_1/a)}{\cosh^2(v_1/a)}
+\frac{\tanh(v_2/a)}{\cosh^2(v_2/a)}
\right],
\label{eq:Udot_reduced}
\end{equation}
with $F=\cosh(\Delta v/a)-\cos(\Delta u)$ and $V=V(\Delta v)$.
Thus the azimuthal drift can be reconstructed by quadrature,
\begin{equation}
U(t)=U(0)+\int_0^t \mathcal{U}\big(\Delta v(s);H,J\big)\,ds,
\end{equation}
where $\mathcal{U}$ denotes the right-hand side of Eq.~\eqref{eq:Udot_reduced}. We verify this prediction by comparing both the instantaneous drift rate $\dot U(t)$ and the reconstructed $U(t)$ with the full numerical solution. As shown in Fig.~\ref{fig:uver}, the reduced theory reproduces both quantities with excellent accuracy over long times. This demonstrates that the observed secular azimuthal drift is not an independent degree of freedom, but arises entirely from the constrained one--dimensional dynamics governed by $(H,J)$.
\subsection{Collective drift of localized vortex clusters}
\label{clusteroutlook}
\begin{figure*}[t]
\centering

\begin{minipage}{0.42\textwidth}
\centering
\includegraphics[width=\linewidth]{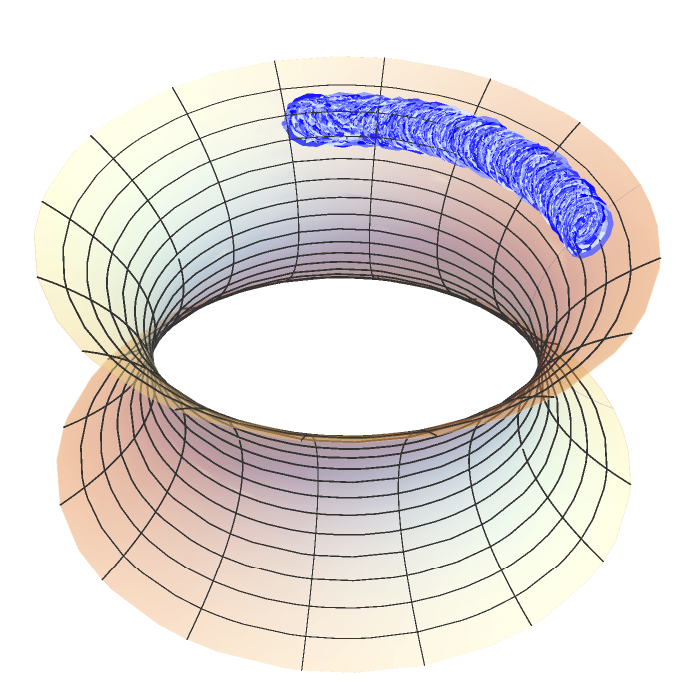}
\end{minipage}
\hfill
\begin{minipage}{0.53\textwidth}
\centering
\includegraphics[width=\linewidth]{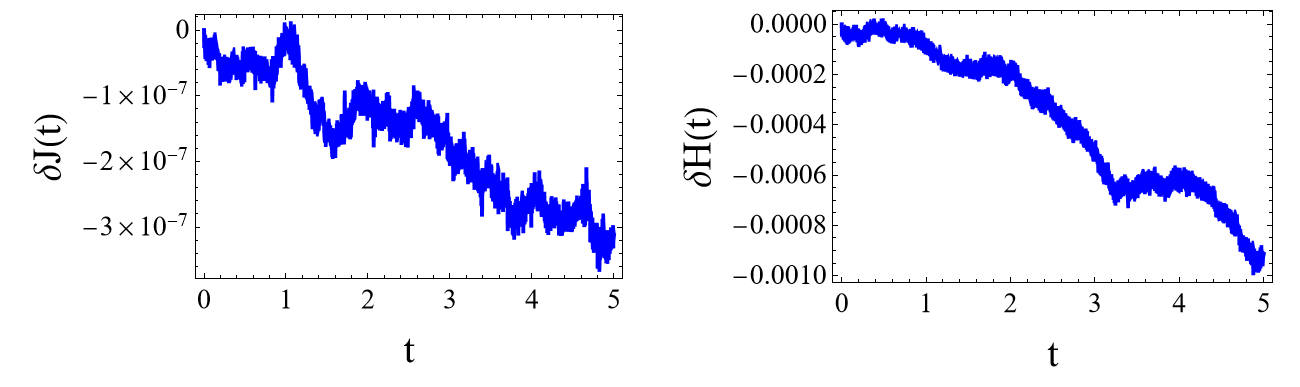}
\end{minipage}

\vspace{0.35cm}

\begin{minipage}{0.48\textwidth}
\centering
\includegraphics[width=\linewidth]{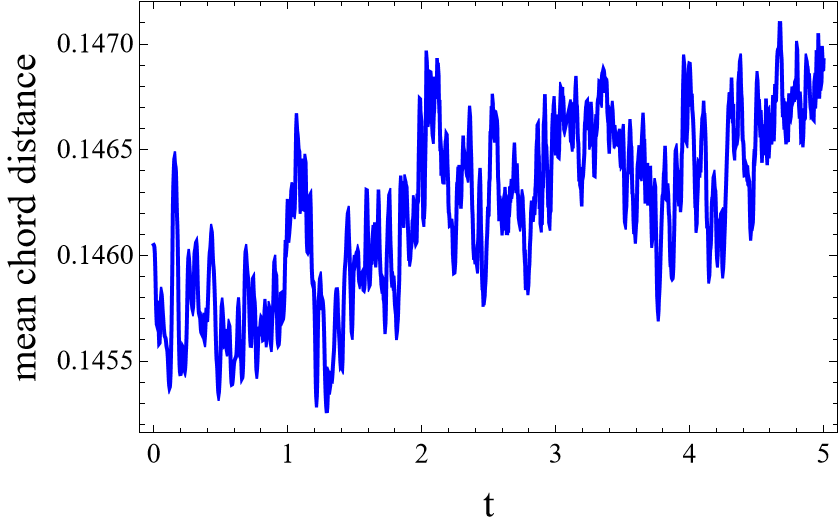}
\end{minipage}
\hfill
\begin{minipage}{0.48\textwidth}
\centering
\includegraphics[width=\linewidth]{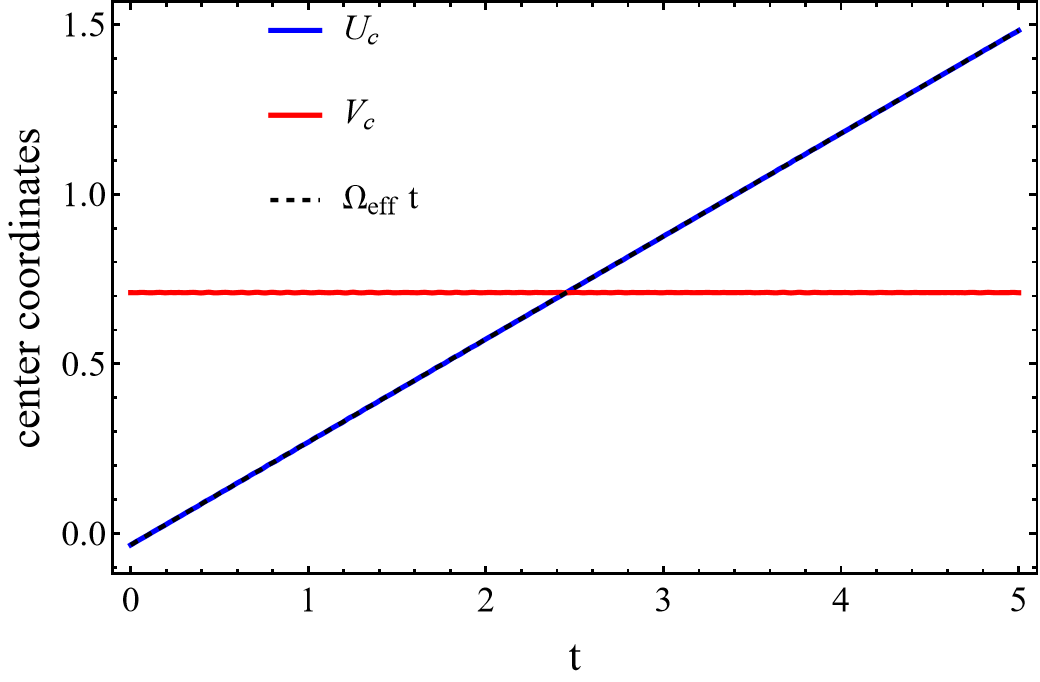}
\end{minipage}

\caption{
Collective many-vortex dynamics on the catenoid. The simulation
uses a localized same-sign cluster of $N=10$ identical vortices with
$\Gamma_i=1$ and $a=1$, initialized near $(U_c,V_c)=(0,0.7)$ with spread
$\epsilon_u=\epsilon_v=0.12$. Top left: embedded trajectories on the catenoid,
showing that the vortices remain grouped as a compact packet while drifting
azimuthally. Top right: conservation diagnostics for the rotational momentum
$J$ and Hamiltonian $H$, whose deviations remain small during the integration.
Bottom left: mean Euclidean chord distance between vortex pairs, indicating
that the cluster remains localized. Bottom right: center coordinates $U_c(t)$
and $V_c(t)$; the meridional center remains nearly fixed, while the azimuthal
center grows approximately linearly and is well described by an effective
drift $\Omega_{\rm eff}t$. This collective drift is the many-vortex analogue
of the curvature-induced azimuthal drift found for co-rotating vortex
pairs. 
}
\label{fig:cluster_diagnostics_main}
\end{figure*}
The two--vortex analysis above shows that curvature generically produces an azimuthal drift even when the relative motion remains bounded. It is therefore natural to ask whether a similar mechanism persists in many-vortex systems. As an illustration, we consider a numerical simulation of a localized same-sign vortex cluster on the catenoid, shown in Fig.~\ref{fig:cluster_diagnostics_main}. The simulation consists of $N=10$ identical vortices initialized in a compact cluster away from the throat, where the curvature-gradient factor
\[
\tanh(V_c/a)\sech^2(V_c/a)
\]
is nonzero.

The numerical trajectories show that the vortices remain localized while the cluster undergoes a coherent azimuthal drift around the catenoid. Defining the cluster center by
\[
U_c(t)=\frac1N\sum_{i=1}^N u_i(t),
\qquad
V_c(t)=\frac1N\sum_{i=1}^N v_i(t),
\]
we find that $V_c(t)$ remains approximately constant whereas $U_c(t)$ grows nearly linearly in time, indicating collective transport along the azimuthal direction. The conservation of the Hamiltonian and rotational momentum remain stable throughout the integration, while the mean inter-vortex separation remains bounded, showing that the cluster does not disperse over the simulated time interval. These results provide  evidence that the curvature-induced drift identified for vortex pairs may persist at the collective level for localized many-vortex states. A systematic reduced theory for such cluster dynamics, including effective drift laws and shape evolution, will be developed in future work \cite{samanta2026}.

\section{Conclusion}
\label{cncl}
We have developed a Hamiltonian framework for the motion of point vortices on a surface of variable negative curvature, using the catenoid geometry. We used it to obtain an explicit analytic reduction of the co-rotating two-vortex problem. The geometry enters the dynamics through both the pairwise interaction kernel and the curvature-induced Robin term (regular part of the Green's function), the latter generating self-induced azimuthal motion absent in the planar problem. Exploiting the rotational symmetry of the catenoid, we identified a conserved momentum in addition to the Hamiltonian and showed that the two-vortex system is Liouville integrable with explicit one dimensional integral solution.

Within this framework, we identified an exact antipodal co-rotating solution for identical vortices and showed that its angular velocity is controlled by the curvature gradient rather than by the curvature itself. In particular, the rotation rate can be expressed directly in terms of the Gaussian curvature profile,
\[
\Omega(V)=
\frac{\Gamma}{16\pi}
\frac{K'(V)}{\sqrt{-K(V)}}.
\]
 The associated linear stability analysis further revealed that the same geometric factor governing the rigid rotation also determines the instability rate of the symmetric state,
\[
\lambda=\sqrt3\,|\Omega(V_0)|.
\]
Beyond the symmetric configuration, the generic equal-strength co-rotating dynamics exhibits bounded relative oscillations together with a secular azimuthal drift. The reduced formulation shows that this drift is not an independent dynamical degree of freedom, but is entirely determined by the conserved quantities and the reduced one-dimensional motion. The numerical integrations confirm the reduced analytic description and azimuthal drift. The many-vortex simulations indicate that the curvature-induced drift persists at the collective level: localized same-sign vortex clusters remain localized while undergoing an effective azimuthal drift around the catenoid. 

At the same time, the present analysis has several limitations. The model is based on the ideal point-vortex approximation, treating vortices as singular objects with negligible core size in an inviscid incompressible fluid on a fixed surface. Consequently, finite-core structure, compressibility, viscous dissipation and coupling to surface deformations are neglected. On curved manifolds, finite-core effects may become important when the vortex core size is comparable to local curvature scales, potentially modifying both the effective interactions and the drift and stability properties derived here.

Dissipative effects such as viscosity, thermal damping, or mutual friction are also expected to break the exact Hamiltonian structure underlying the present formulation, leading to secular energy loss and altered long-time dynamics, see for example Ref.~\cite{aswathy2026b}. In addition, the explicit integrability established here relies strongly on the axial symmetry of the catenoid and the two-vortex case, and is not expected to persist generically for arbitrary curved surfaces or large vortex ensembles. Finally, the many-vortex results presented here remain exploratory and do not yet constitute a systematic reduced theory for collective drift or shape dynamics.

Several extensions naturally follow from this work. It would be interesting to determine whether analogous curvature-gradient-driven motion arise on more general negatively curved surfaces, including geometries without rotational symmetry. Another important direction is the development of reduced descriptions for localized vortex clusters, where collective drift and curvature-induced interactions may compete. Extensions to finite-core vortices, dissipative systems, externally driven flows, and quantum-fluid realizations may also help clarify the robustness of the dynamical features identified here. Taken together, the results of this work show how curvature, symmetry, and vortex interactions combine to produce analytically tractable nontrivial dynamics.

\section{Acknowledgements}
It is a pleasure to thank Suryateja Gavva, Naomi Oppenheimer and Haim Diamant. R.S is supported by DST INSPIRE Faculty fellowship, India (Grant No.IFA19-PH231). Both authors acknowledge support from NFSG and OPERA Research Grant from Birla Institute of Technology and Science, Pilani (Hyderabad Campus). 

\section*{Data Availability}

All data generated or analyzed during this study are included in this
article. The computational codes used in this work are available from
the corresponding author upon reasonable request.
\appendix

\section{Liouville integrability of the two--vortex system on the catenoid}
\label{app1}
In this appendix we establish explicitly that the two--vortex problem on the catenoid is Liouville integrable. The proof proceeds by identifying the symplectic structure, constructing the associated Poisson bracket, and demonstrating that the Hamiltonian and rotational momentum are conserved quantities in involution. For two vortices, the phase space is four-dimensional, with coordinates
\begin{align}
(v_1,u_1,v_2,u_2).\nn
\end{align}
Defining \(h(v)=\cosh(v/a)\), the dynamics is formulated on the
four-dimensional phase space \((v_1,u_1,v_2,u_2)\) equipped with the
symplectic form
\begin{align}
\omega
=
\sum_{i=1}^2
\Gamma_i a h^2(v_i)\,
du_i\wedge dv_i .
\label{sympform}
\end{align}
We will be adopting the convention
\begin{align}
\iota_{X_H}\omega=-dH,
\label{HamConvention}
\end{align}
where \(X_H\) is the Hamiltonian vector field,
\begin{align}
X_H
=
\sum_{i=1}^2
\left(
\dot u_i\,\frac{\partial}{\partial u_i}
+
\dot v_i\,\frac{\partial}{\partial v_i}
\right).
\end{align}
Using
\begin{align}
\iota_{X_H}
\left(
du_i\wedge dv_i
\right)
=
\dot u_i\,dv_i
-
\dot v_i\,du_i,
\end{align}
the relation \eqref{HamConvention} gives the Hamilton's equations of the main text,
\begin{align}
\Gamma_i a h^2(v_i)\,\dot v_i
&=
\frac{\partial H}{\partial u_i},
\\
\Gamma_i a h^2(v_i)\,\dot u_i
&=
-
\frac{\partial H}{\partial v_i},
\qquad i=1,2.
\label{HamEqsAppendix}
\end{align}
The inverse of the symplectic form determines the associated Poisson structure. For two phase-space functions
\(
A(v_1,u_1,v_2,u_2)
\)
and
\(
B(v_1,u_1,v_2,u_2),
\)
the Poisson bracket is
\begin{align}
\{A,B\}
=
\sum_{i=1}^2
\frac{1}{\Gamma_i a h^2(v_i)}
\left(
\frac{\partial A}{\partial v_i}
\frac{\partial B}{\partial u_i}
-
\frac{\partial A}{\partial u_i}
\frac{\partial B}{\partial v_i}
\right).
\label{PBdef}
\end{align}
In particular, the fundamental brackets are
\begin{align}
\{v_i,u_j\}
&=
\frac{\delta_{ij}}
{\Gamma_i a h^2(v_i)},\nn
\\
\{v_i,v_j\}
&=
0,\nn
\\
\{u_i,u_j\}
&=
0.\nn
\label{FundamentalPB}
\end{align}
The Hamiltonian evolution of any phase-space function \(F\) is therefore
\begin{align}
\dot F=\{F,H\},
\end{align}
which reproduces the equations of motion \eqref{HamEqsAppendix}.
The Hamiltonian for the two--vortex system is
\begin{align}
H
&=
\frac{\Gamma_1\Gamma_2}{4\pi}
\log\!\left[
\cosh\!\left(\frac{v_1-v_2}{a}\right)
-
\cos(u_1-u_2)
\right] -
\frac{1}{4\pi}
\left[
\Gamma_1^2\log h(v_1)
+
\Gamma_2^2\log h(v_2)
\right].\nn
\label{Happendix}
\end{align}
The first term is the geometric generalization of the planar logarithmic interaction energy, while the second term is the curvature-induced self-energy arising from the nonuniform geometry of the catenoid. The conserved angular momentum is
\begin{align}
J
&=
\Gamma_1
\left[
\frac{a}{2}v_1
+
\frac{a^2}{4}\sinh\!\left(\frac{2v_1}{a}\right)
\right]+
\Gamma_2
\left[
\frac{a}{2}v_2
+
\frac{a^2}{4}\sinh\!\left(\frac{2v_2}{a}\right)
\right].
\end{align}
Equivalently, writing
\begin{align}
S(v)
=
\frac{a}{2}v
+
\frac{a^2}{4}\sinh\!\left(\frac{2v}{a}\right),
\qquad
J=\sum_{i=1}^2\Gamma_i S(v_i),
\end{align}
one has
\begin{align}
S'(v)=a\cosh^2(v/a)=a h^2(v).
\end{align}
We now show that \(J\) generates the rotational symmetry. Since \(J\) depends only on \(v_1,v_2\),
\begin{align}
\frac{\partial J}{\partial u_i}=0,
\qquad
\frac{\partial J}{\partial v_i}
=
\Gamma_i S'(v_i)
=
\Gamma_i a h^2(v_i).
\label{dJ}
\end{align}
Using the Poisson bracket \eqref{PBdef},
\begin{align}
\{u_i,J\}
&=
\sum_{k=1}^2
\frac{1}{\Gamma_k a h^2(v_k)}
\left(
\frac{\partial u_i}{\partial v_k}
\frac{\partial J}{\partial u_k}
-
\frac{\partial u_i}{\partial u_k}
\frac{\partial J}{\partial v_k}
\right)
\nonumber\\
&=
-
\frac{1}{\Gamma_i a h^2(v_i)}
\frac{\partial J}{\partial v_i}
\nonumber\\
&=
-1,
\label{uJ}
\end{align}
while
\begin{align}
\{v_i,J\}=0.
\label{vJ}
\end{align}
We next establish that \(H\) and \(J\) are in involution. Since the Hamiltonian depends on \(u_1,u_2\) only through the difference
\begin{align}
\Delta u=u_1-u_2,\nn
\end{align}
it follows that
\begin{align}
\frac{\partial H}{\partial u_1}
+
\frac{\partial H}{\partial u_2}
=
0.
\label{usym}
\end{align}
Using \eqref{PBdef} and \eqref{dJ},
\begin{align}
\{H,J\}
&=
\sum_{i=1}^2
\frac{1}{\Gamma_i a h^2(v_i)}
\left(
\frac{\partial H}{\partial v_i}
\frac{\partial J}{\partial u_i}
-
\frac{\partial H}{\partial u_i}
\frac{\partial J}{\partial v_i}
\right)
\nonumber\\
&=
-
\sum_{i=1}^2
\frac{1}{\Gamma_i a h^2(v_i)}
\frac{\partial H}{\partial u_i}
\Gamma_i a h^2(v_i)
\nonumber\\
&=
-
\left(
\frac{\partial H}{\partial u_1}
+
\frac{\partial H}{\partial u_2}
\right)
\nonumber\\
&=
0.
\label{HJzero}
\end{align}
Thus the Hamiltonian and rotational momentum are conserved quantities in involution:
\begin{align}
\{H,J\}=0.
\end{align}
The two--vortex phase space is four-dimensional and therefore corresponds to a Hamiltonian system with two degrees of freedom. Liouville integrability requires two independent conserved quantities in involution. The pair \((H,J)\) provides precisely these conserved quantities. Away from collision configurations, the differentials \(dH\) and \(dJ\) are functionally independent, and therefore the system is Liouville integrable.


\begin{thebibliography}{99}

\bibitem{aref}
H. Aref,
\emph{Integrable, chaotic, and turbulent vortex motion in two-dimensional flows},
\emph{Annu.\ Rev.\ Fluid Mech.} {\bf 15} (1983) 345--389.

\bibitem{saffman}
P.\,G. Saffman,
\emph{Vortex Dynamics}
(Cambridge Univ.\ Press, 1993).

\bibitem{lin1}
C.\,C. Lin,
\emph{On the motion of vortices in two dimensions: I. Existence of the Kirchhoff--Routh function},
\emph{Proc.\ Natl.\ Acad.\ Sci.\ USA} {\bf 27} (1941) 570--575.

\bibitem{lin2}
C.\,C. Lin,
\emph{On the motion of vortices in two dimensions: II. Some further investigations on the Kirchhoff--Routh function},
\emph{Proc.\ Natl.\ Acad.\ Sci.\ USA} {\bf 27} (1941) 575--577.
\bibitem{Tchieu2012}
A.~A.~Tchieu, E.~Kanso, and P.~K.~Newton,
\textit{The finite-dipole dynamical system},
Proc.\ R.\ Soc.\ A \textbf{468}, 3006--3026 (2012).
\bibitem{Lydon2022}
K.~Lydon, S.~V.~Nazarenko, and J.~Laurie,
\emph{Dipole dynamics in the point vortex model},
J.\ Phys.\ A: Math.\ Theor. \textbf{55}, 385702 (2022).
\bibitem{sam1}
R. Samanta and N. Oppenheimer,
\emph{Vortex flows and streamline topology in curved biological membranes},
\emph{Phys.\ Fluids} {\bf 33} (2021) 092111.

\bibitem{sam2}
U. Maurya, S.\,T. Gavva, A. Saha and R. Samanta,
\emph{Vortex dynamics in tubular fluid membranes},
\emph{Phys.\ Fluids} {\bf 37} (2025) 073109.

\bibitem{sam3}
Aswathy K.\,R. , U. Maurya, S.\,T. Gavva and R. Samanta,
\emph{Dynamics of vortex clusters on a torus},
\emph{Phys.\ Fluids} {\bf 37} (2025) 093324.
\bibitem{banthia2026}
K.~Banthia and R.~Samanta,
\emph{A self propelled vortex dipole model on a surface of variable negative curvature},
J.\ Phys.\ A: Math.\ Theor.\ \textbf{59}, 145701 (2026).

\bibitem{aswathy2026}
 Aswathy K.\,R. and R. Samanta,
\emph{Collective dynamics of vortex clusters in compact fluid domains: From pair interactions to a quadrupole description},
arXiv:2604.07373 [physics.flu-dyn] (2026).



\bibitem{bg}
V.\,A. Bogomolov,
\emph{Dynamics of vorticity on a sphere},
\emph{Fluid Dyn.} {\bf 12} (1977) 863--870.
\bibitem{kimok}
Y. Kimura and H. Okamoto,
\emph{Vortex motion on a sphere},
\emph{J.\ Phys.\ Soc.\ Jpn.} {\bf 56} (1987) 4203--4206.

\bibitem{kimura}
Y. Kimura,
\emph{Vortex motion on surfaces with constant curvature},
\emph{Proc.\ R.\ Soc.\ A} {\bf 455} (1999) 245--259.

\bibitem{newton1}
R. Kidambi and P.\,K. Newton,
\emph{Streamline topologies for integrable vortex motion on a sphere},
\emph{Physica D} {\bf 140} (2000) 95--125.
\bibitem{newton2}
R. Kidambi and P.\,K. Newton,
\emph{Point vortex motion on a sphere with solid boundaries},
\emph{Phys.\ Fluids} {\bf 12} (2000) 581--588.
\bibitem{hally}
D. Hally,
\emph{Stability of streets of vortices on surfaces of revolution with a reflection symmetry},
\emph{J.\ Math.\ Phys.} {\bf 21} (1980) 211--217.




\bibitem{crowdymarshall}
D.G. Crowdy and J. Marshall,
\emph{Analytical formulae for the Kirchhoff--Routh path function in multiply connected domains},
\emph{Proc.\ R.\ Soc.\ A} {\bf 461} (2005) 2477--2501.

\bibitem{boattok}
S.~Boatto and J.~Koiller,
\emph{Vortices on closed surfaces},
in \emph{Geometry, Mechanics, and Dynamics: The Legacy of Jerry Marsden},
edited by D.~E.~Chang, D.~D.~Holm, G.~Patrick, and T.~Ratiu,
Fields Institute Communications, Vol.~73
(Springer, New York, 2015), pp.~185--237.


\bibitem{Koiller2009}
J.~Koiller and S.~Boatto,
\emph{Vortex pairs on surfaces},
AIP Conf. Proc., \textbf{1130}, 77--88 (2009).
\bibitem{boattod}
D.\,G. Dritschel and S. Boatto,
\emph{The motion of point vortices on closed surfaces},
\emph{Proc.\ R.\ Soc.\ A} {\bf 471} (2015) 20140890.


\bibitem{Turner2010}
A.\,M. Turner, V. Vitelli and D.\,R. Nelson,
\emph{Vortices on curved surfaces},
\emph{Rev.\ Mod.\ Phys.} {\bf 82} (2010) 1301--1348.


\bibitem{voigt}
S.~Reuther and A.~Voigt,
\emph{The interplay of curvature and vortices in flow on curved surfaces},
\emph{Multiscale Model.\ Simul.} \textbf{13}, 632--643 (2015).





\bibitem{Gustafsson2022}
B. Gustafsson,
\emph{Vortex pairs and dipoles on closed surfaces},
\emph{J.\ Nonlinear Sci.} {\bf 32} (2022) 62.



\bibitem{khesin2024}
T.~D.~Drivas, D.~Glukhovskiy, and B.~Khesin,
\emph{Singular Vortex Pairs Follow Magnetic Geodesics},
Int.\ Math.\ Res.\ Not.\ \textbf{2024}(14), 10880--10894 (2024).



\bibitem{Caracanhas2022}
M.~A.~Caracanhas, P.~Massignan, and A.~L.~Fetter,
\emph{Superfluid vortex dynamics on an ellipsoid and other surfaces of revolution},
Phys.\ Rev.\ A \textbf{105}, 023307 (2022).

\bibitem{Neely2010}
T.\,W. Neely, E.\,C. Samson, A.\,S. Bradley, M.\,J. Davis and B.\,P. Anderson,
\emph{Observation of vortex dipoles in an oblate Bose--Einstein condensate},
\emph{Phys.\ Rev.\ Lett.} {\bf 104} (2010) 160401.

\bibitem{Freilich2010}
D.\,V. Freilich, D.\,M. Bianchi, A.\,M. Kaufman, T.\,K. Langin and D.\,S. Hall,
\emph{Real-time dynamics of single vortex lines and vortex dipoles in a Bose--Einstein condensate},
\emph{Science} {\bf 329} (2010) 1182--1185.

\bibitem{vsc}
G.~Gauthier, M.~T.~Reeves, X.~Yu, A.~S.~Bradley, M.~Baker, T.~A.~Bell, 
H.~Rubinsztein-Dunlop, M.~J.~Davis, and T.~W.~Neely,
\textit{Giant vortex clusters in a two-dimensional quantum fluid},
Science \textbf{364}, 1264--1267 (2019).

\bibitem{Rooney2011}
S.\,J. Rooney, P.\,B. Blakie, B.\,P. Anderson and A.\,S. Bradley,
\emph{Suppression of Kelvon-induced decay of quantized vortices in oblate Bose-Einstein condensates},
\emph{Phys.\ Rev.\ A} {\bf 84}, 023637 (2011). 

\bibitem{Goodman2015}
R.\,H. Goodman, P.\,G. Kevrekidis and R. Carretero-Gonz\'alez,
\emph{Dynamics of Vortex Dipoles in Anisotropic Bose–Einstein Condensates},
\emph{SIAM J.\ Appl.\ Dyn.\ Syst.} {\bf 14}, no. 2,  699--729 (2015).

\bibitem{white2012}
A.\,C. White, C.\,F. Barenghi and N.\,P. Proukakis,
\emph{Creation and Characterization of Vortex Clusters in Atomic Bose-Einstein Condensates},
\emph{Physical Review A} {\bf 86}, 013635, (2012).

\bibitem{White2014}
A.~C.~White, B.~P.~Anderson, and V.~S.~Bagnato,
\textit{Vortices and turbulence in trapped atomic condensates},
Proc.\ Natl.\ Acad.\ Sci.\ U.S.A.\ \textbf{111}, 4719--4726 (2014).


\bibitem{Stagg2016}
G.~W.~Stagg, N.~G.~Parker, and C.~F.~Barenghi, \emph{Ultraquantum turbulence in a quenched homogeneous Bose gas}, Phys.\ Rev.\ A \textbf{94}, 053632 (2016).
\bibitem{samanta2026}
R.~Samanta,
\emph{Co-rotating Vortex Clusters on Negatively Curved Geometries},
manuscript in preparation (2026).

\bibitem{aswathy2026b}
Aswathy~KR and R.~Samanta,
\textit{Dissipative Vortex Binaries in Compact Fluid Domains with Geometric Corrections},
arXiv:2604.23857 [physics.flu-dyn] (2026).

\end{thebibliography}
\end{document}